Hometronics – Accessible production of graphene suspensions for health sensing applications using only household items

Adel K.A. Aljarid[†,1], Jasper Winder[†,1], Cencen Wei[†,1], Arvind Venkatraman[2], Oliver Tomes[3], Aaron Soul[3], Dimitrios G. Papageorgiou[3], Matthias E. Möbius[2] and Conor S. Boland[*,1]

[1]*School of Mathematical and Physical Sciences, University of Sussex, Brighton, BN1 9QH, U.K.*

[2]*School of Physics, AMBER and CRANN Research Centres, Trinity College Dublin, Dublin 2, Ireland.*

[3]*School of Engineering and Materials Science, Queen Mary University, London, E1 4NS, U.K.*

[†]Authors contributed equally to the work

*Corresponding Author. Email: c.s.boland@sussex.ac.uk (Conor S. Boland)

**Abstract**

Nanoscience at times can seem out of reach to the developing world and the general public, with much of the equipment expensive and knowledge seemingly esoteric to nonexperts. Using only cheap, everyday household items, accessible research with real applications can be shown. Here, graphene suspensions were produced using pencil lead, tap water, kitchen appliances, soaps and coffee filters, with a children's glue-based graphene nanocomposite for highly sensitive pulse measurements demonstrated.

Research into using liquid exfoliation (LE) methods have led to the wide-spread production of two-dimensional (2D) materials on a large scale.[1] Particularly, the demonstration of shear exfoliation methods[2] has made LE more viable for low-cost production via the use of household food processors.[3,4] However, low-income institutions, third world countries and citizen science initiatives would still struggle. As beyond the shearing device, the reagents and purifying instruments (i.e. centrifuge) commonly applied are still economically exclusive.[5] Here, we demonstrate that a pencil can be processed into a powder and LE in a blender to isolate a suspension of high-quality graphene in tap water using a coffee filter (Figure 1a). Furthermore, we show that by using school glue, a nanocomposite we call pencil putty (p-putty) can be produced for impactful health applications. The total overall costing of this project was a mere £76.22 (equivalent to ~$96.67, Table S1), greatly facilitating its accessibility.

In Figures 1b – e , we show the pencil and blender (with graphene suspension contained inside before and after blending) utilised in the study, as well as the stable graphene suspension produced. For the exfoliation medium, a mixture of dish soap in tap water was used. Fabric softener was also added to the blend to supress foaming. After several exfoliation intervals (see Methods), the resultant suspension was passed through a coffee filter with a pore diameter of ~10 µm. This resulted in the collection of large particulates of unexfoliated graphite on the coffee filter, with the suspension that passed through gathered in a flask. Using atomic force microscopy (AFM) on the filtered suspension, the pencil lead was confirmed to have been exfoliated into nanosheets in Figure 1d. Through Raman spectroscopy in Figure 1e, the production of graphene was confirmed through a comparative analysis of the powdered and exfoliated pencil lead spectra. All characteristic modes associated with graphitic material were present in both samples at Raman shifts of $\sim 1350$ cm$^{-1}$, $\sim 1580$ cm$^{-1}$, and $\sim 2700$ cm$^{-1}$, which relate to the D, G and 2D modes respectively.[6] Specifically looking at the 2D modes of the samples, there is a downshift from the powdered (~2705 cm$^{-1}$) to the exfoliated (~2697 cm$^{-1}$) material, indicative of LE leading to multi-layered graphene.[7] Using standardised spectroscopic metrics,[8] information with regards to the graphene's mean length (<*L*>) in µm,

$$\langle L \rangle = \frac{0.094}{(I_D/I_G)_{Graphene} - (I_D/I_G)_{Graphite}} \tag{1}$$

And mean layer number (<*N*>),

$$\langle N \rangle = 1.04 \cdot \left[(I_{2D}/I_G)_{Graphene}\right]^{-2.32} \tag{2}$$

can be derived from the Raman spectra. Where ($I_D$/$I_G$)$_{Graphene}$ and ($I_D$/$I_G$)$_{Graphite}$, are the ratios between the D and G mode intensities for graphene and the starting graphite source respectively. Furthermore, ($I_{2D}$/$I_G$)$_{Graphene}$ is the ratio between the 2D and G mode intensities. Using Equations 1 and 2, for our home-made graphene, <*L*> ~ 314 ± 63 nm and <*N*> ~ 7.37 ± 1.84 respectively. In comparison to standard lab processed graphene, similar aspect ratio values were reported.[9]

In addition to demonstrating purely kitchen appliance-based graphene production, we also show that the resultant nanosheets can be applied in an equally accessible application with powerful research outcomes. Using a commercial water-based polyvinyl alcohol (PVA) children's glue kit, graphene could be added via solution mixing to create p-putty (Figure 2a). This nanocomposite was based on a boron crosslinked hydrogel created from mixing Elmer's PVA glue with a boric acid solution known as Elmer's Magic Liquid. We found that the addition of the graphene to pristine putty darkened the hue of the material, with the presence of graphene also turning the putty from an insulator to an electrical conductor (Figure S1). Through scanning electron microscopy (SEM), the topography between the pristine (Figure 2b) and a graphene loaded sample (Figure 2c) are also seen to greatly differ. Mechanically, the addition of graphene to the putty had no effect on either the Young's modulus ($Y$) in Figure 2d nor the other mechanical properties (Figures S2 and S3) of the nanocomposites as a function of volume fraction ($V_f$). This invariance in mechanical properties is a common occurrence when pristine graphene was mixed into a hydrogel matrix due to its hydrophobic nature.[9,10] In the case of p-putty, it resulted in a material that remained very pliable (Movie S1). However, it was noted that at higher loadings ($V_f \sim 0.6\%$), visible graphene aggregation occurred in the samples, which resulted in mechanical properties to marginally increase. Through rheological measurements in Figure 2e, a 0.4% p-putty sample behaved similarly to the unloaded equivalent as a function of shear strain. Both materials presented Maxwell fluid behaviour, with a relaxation time of ~2 s and a viscosity ~3000 Pa·s (Figure S4).

Upon the application of compressive strain ($-\varepsilon$), the electrical properties of p-putty were observed to change. In Figure 2f, fractional resistance change ($\Delta R/R_0$) as a function of $-\varepsilon$ for a select range of $V_f$ increased linearly in accordance with electromechanical theory.[11] Using the following expression

$$G = \frac{\Delta R}{R_0} \cdot \varepsilon^{-1} \qquad (3)$$

The electromechanical sensitivity ($G$) can be extrapolated by fitting Equation 3 to the low strain regimes of $\Delta R/R_0$ versus $-\varepsilon$ curves (Figure S5), with $G$ values plotted as a function of $V_f$ in Figure 2g. As standard, $G$ generally is a maximum near the percolation threshold due to minimised network connections and decreases with $V_f$ as connections increase.[12] We however note that due to common network necking effects,[10,13] which leads to inefficient strain transfer to the network during deformation, $G$ increased from ~27 at ~0.1% to ~57 at ~0.4% and subsequently decreased to ~45 at ~0.6%. In comparison to other hydrogel[10] and mixed-phase[11] nanocomposite strain sensors, $G$ is generally < 50.[14] Through Figure S5, another important performance metric which dictates the value of $G$, can also be derived from the strain limit of the linear regime. Working factor ($W$) was plotted versus $V_f$ in Figure 2h, alongside the yield strain ($\varepsilon_Y$). As expected,[11] values for $W$ and $\varepsilon_Y$ were found to overlay one another due to their intrinsic connection. In Figure 2i, $G$ as a function of $W$ showed that in the necking regime $W$ was invariant.[10] However, in the percolation regime, $W$ scaled with an exponent of $-m$.[9,14] Where $m$ is the Kraus constant and has a value of 0.5. With p-putty's combination of exceedingly soft stiffness and large electromechanical sensitivity, the nanocomposite was capable of measuring heart rate as a function of time (Figure 2j). When held to the carotid artery in the neck, we report a steady state signal and through fast Fourier transform (FFT) analysis a healthy pulse frequency of ~1.2 Hz (equivalent to 72 bpm).

In this communication, we introduce hometronics as not only a valid way in which nanoscience can be accessible to all researchers and the public alike, but a simple methodology with real applications from our everyday surroundings. Nanoscience research need not be complex and by taking pencil lead processed using home appliances, we demonstrate graphene suspensions from household detergents. Using home-based graphene, nanocomposites from children's glue can be produced to easily measure a wearer's pulse. The accessibility of the work also leads to the use of the methodologies being applied in schools with students from a range of economic backgrounds to inspire and diversify the next generation of scientists.

**Methods**

Graphene Exfoliation

A Graphite Pure 2900 9B pencil was initially crushed in a mortar and pestle, then ground in a SQ Professional Blitz Coffee Grinder - Spice Grinding Mill - One-Touch for ~60 s. A mass of 2.5 g of Fairy Max Power Washing Up Liquid was then mixed with 1 L of tap water. 300 mL of the soap mixture was then added to a Tefal BL420840 Blendforce II Blender (Plastic Jug, 600 W, 1.5 L) with 1 mL of Ecover Fabric Softener also added to the blender. 16 g of powdered pencil lead was then placed in the blender and stirred into the soap/softener/water mixture by hand. The blender was then turned on to the first (#1) setting for 1-minute on, 1 minute off intervals to reduce motor heating effects, until the total blending time reached 15 minutes. The solution was then transferred to a beaker and left to sit for 24 hrs. After which, the solution was passed through a Rombouts Italian No. 2 Coffee Filter.

Atomic Force Microscopy

Diluted graphene suspension (~ 0.01 mg/mL) was drop cast onto a heated silicon wafer (70 °C). The wafer was then washed with deionized water to remove residual household soaps. A Dimension Icon Bruker in an insulated box over an anti-vibrant stage to minimize environmental noise and building vibrations was used for measurement. A ScanAsyst Air tip with a spring constant of 0.4 N m$^{-1}$ and a tip-sample contact force of 5.0 nN was used.

Raman Spectroscopy

Graphene suspension (6 mg/mL) was drop cast onto a heated silicon wafer (70 °C). The wafer was then washed with deionized water to remove residual household soaps. For the graphite sample, pencil lead was crushed using the mortar and pestle and patted down onto a silicon wafer. Renishaw inVia confocal Raman microscope with 0.8 cm−1 spectral resolution and a 532 nm laser (type: solid state, model: RL53250) was used for measurements. A 2400 mm$^{-1}$ grating at 100× magnification and 5 mW laser power was applied. Each sample curve is an average of ten spectra.

P-Putty Production

For the graphene sample preparation, 2 mL of Elmer's White PVA Glue, 1 mL of Elmer's Glue Slime Magical Liquid Solution, and 0.5 mL of graphene suspension were mixed by hand for approximately 5 to 7 minutes. To vary the graphene loading, different concentration suspensions, ranging from 1 mg/mL to 10 mg/mL, were used. For the pristine putty sample, in place of the graphene suspension, 0.5 mL of tap water was used. For all p-putty samples, they were left at room temperature for 40 minutes before testing.

Scanning Electron Microscopy

The fracture surfaces of the nanocomposites were observed using scanning electron microscopy (SEM, FEI Inspect-F, The Netherlands) with an acceleration voltage of 5 kV. The nanocomposites were cryo-fractured after immersion in liquid $N_2$ and carbon coated with a ≈5 nm layer.

Electro-Mechanical Measurements

A Stable Micro Systems TA-TXplus texture analyser with two insulating plates was used. Electrically conductive copper tape was attached to the inside of their faces. The copper tape contacts were then attached to a Keithley 2614B source meter using silver wire to supply a current across the samples. P-putty samples cyclical in shape (diameter = 5.5 mm and length 6 mm) were placed onto the bottom stationary plate's copper tape contact, with the test arm lowering the second plate's copper tape contact into contact with the top of the sample. Through this setup, mechanical (texture analyser) and electrical (source meter) data could be simultaneously recorded. Samples were deformed at a rate of 6 mm/min.

Rheology

For measurements, an Anton Paar MCR 302e rheometer was used. Samples was flattened prior to lowering the plates (25mm diameter). Gap size was 1.1mm, with normal force quickly settling to zero after reaching gap height. Samples were subsequently allowed to relax for 30 minutes while oscillating the plate at low amplitude (0.1% strain amplitude at 1 Hz).

Pulse Measurements

An electronic patch made out of Ecoflex 00–30 using a previously described method was used to encapsulate the putty sample for testing.[9] For pulse testing it was performed by Adel K.A. Aljarid on his own person by holding the electronic patch to the carotid artery in the neck.

**Data availability**

The authors declare that all data supporting the findings are available within the communication and/or its supplementary information.

**Supplementary Information**

Supporting Information is available online and includes Table S1, Figures S1 to S5, and Movie S1.

**Acknowledgements**

C.W. and C.S.B. acknowledge funding from University of Sussex Strategic Development Fund. A.K.A.A. acknowledges funding through the Saudi Arabian Cultural Bureau. J.W. and C.S.B. acknowledge funding via the University of Sussex's 2023 Junior Research Associate Award Scheme. A.V. and M.E.M. acknowledge funding from Science Foundation Ireland grant SFI 17/CDA/4704.

**Author Contributions**

C.S.B. conceived and designed experiments. J.W. created all graphene suspensions for the study. C.W. performed AFM and Raman on the graphene suspensions. A.K.A.A. made the composite materials and performed electromechanical tests. O.T. performed SEM. A.V. performed rheological testing. C.S.B. analysed and modelled data. D.G.P., M.E.M., C.S.B. contributed materials, tools, and supervision. C.S.B wrote and revised the paper. A.K.A.A.,

C.W., and J.W. contributed equally to the work and may present themselves as the first author on their respective curriculum vitae.

**Figures**

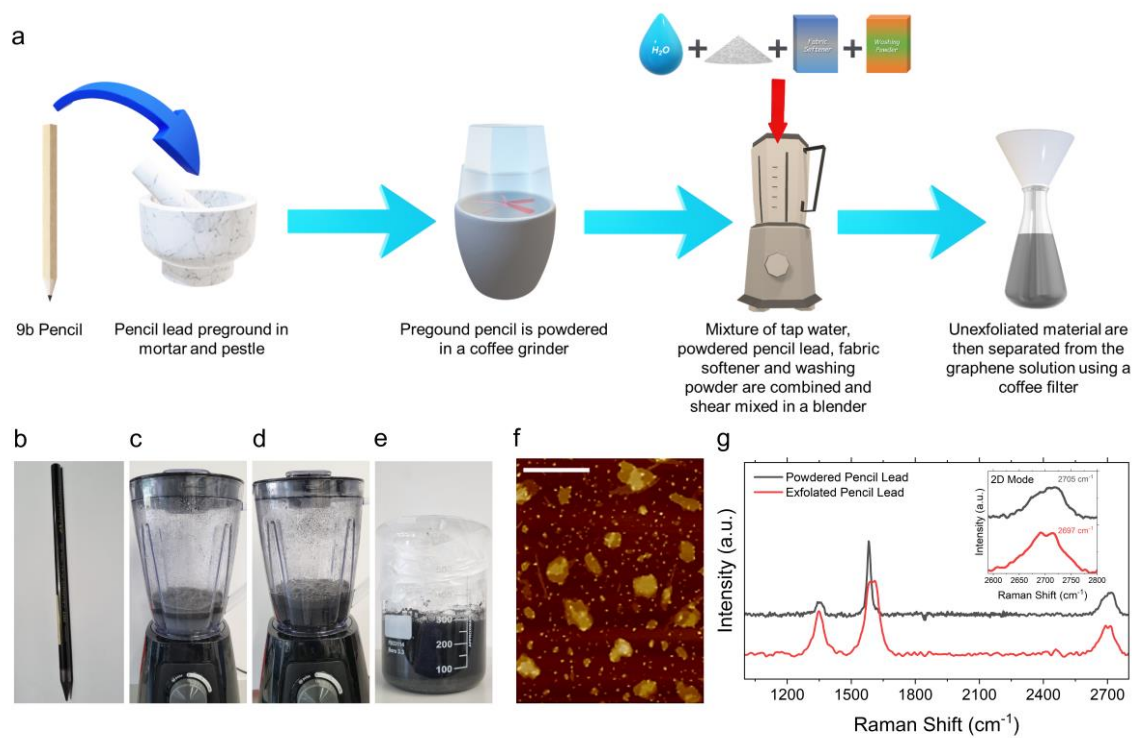

**Figure 1:** Graphene production from household commercial items. **a** Scheme presenting graphene LPE using a pencil. **b** – **e** Photographs of the 9b pencil (**b**), commercial blender before (**c**) and after (**d**) blending and a graphene suspension after sitting for 24 hrs (**e**). **f** Representative AFM micrograph of pencil-based graphene nanosheets. Scale bar is 1000 nm. **g** Raman spectroscopy of powdered and exfoliated pencil lead. Inset, closer look at 2D modes of the respective spectra.

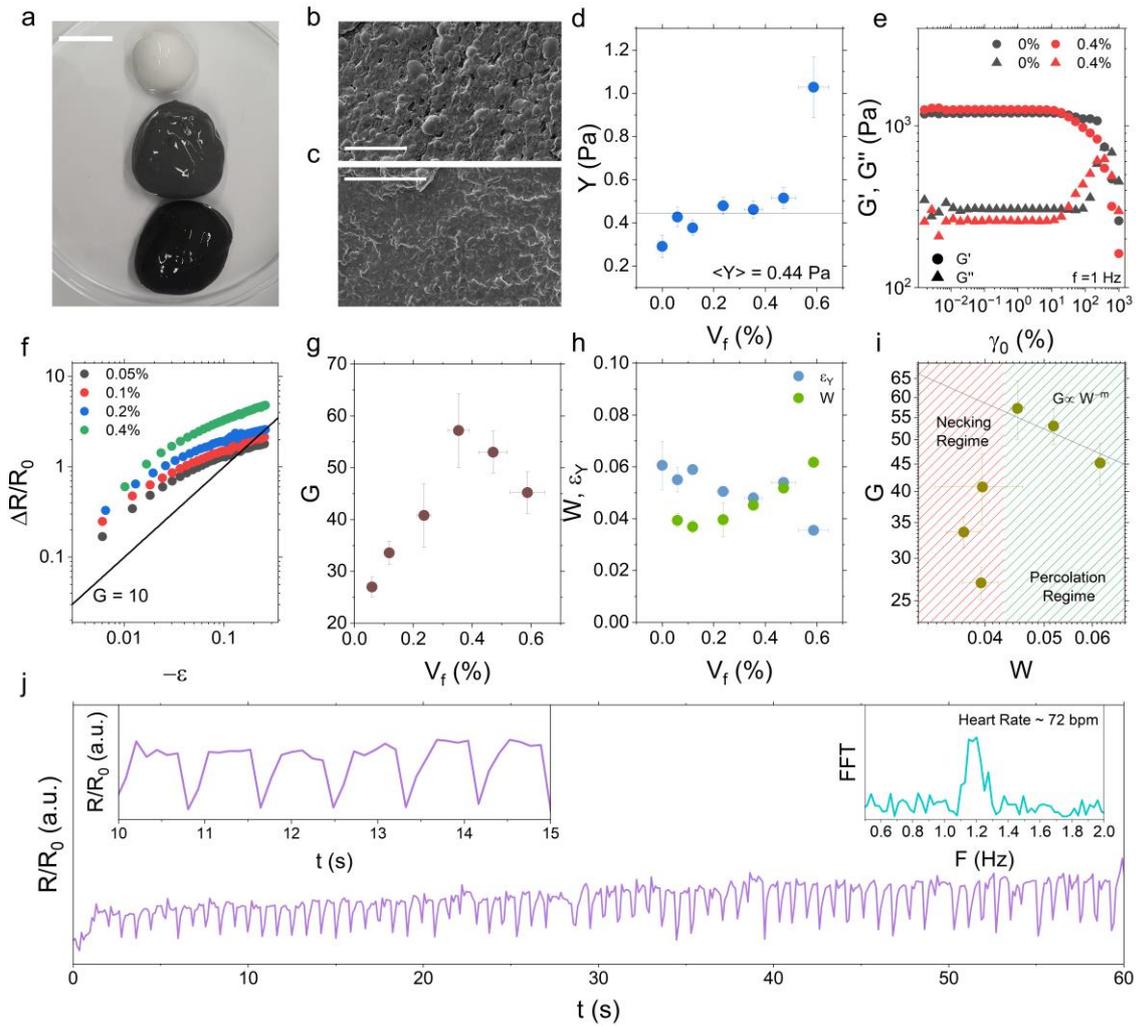

**Figure 2:** P-putty nanocomposite. **a** Photograph of p-putty with 0%, 0.1%, and 0.4% graphene volume fraction ($V_f$), top to bottom. Scale bar is 2 cm. **b, c** SEM micrographs of pristine putty (**b**) and 0.4% p-putty (**c**). Scale bars are 10 µm and 25 µm respectively. **d** Compressive Young's Modulus ($Y$) as a function of $V_f$. Solid line represents mean value. **e** Shear storage ($G'$) and loss ($G''$) moduli of pristine and 0.4% putty versus shear strain amplitude ($\gamma_0$). **f** Fractional resistance change ($\Delta R/R_0$) as a function of compression strain for a select range of $V_f$. Solid line is a fit of Equation 3, where gauge factor ($G$) = 10. **g** P-putty gauge factor ($G$) as a function of $V_f$. **h** P-putty working factor ($W$) and yield strain versus $V_f$. **i** P-putty gauge ($G$) and working ($W$) factors potted against one another scaled with a power-law exponent associated with the Kraus constant ($m$) in the percolation regime. **j** Pulse as a function of time. Insets, zoomed in view of pulse signal (lefthand side) and fast Fourier transform (FFT) of pulse data showing a rate of ~1.2 Hz (righthand side).